\documentclass[10pt]{article}
\usepackage{geometry}
\usepackage{xcolor}
\definecolor{graycolor}{gray}{0.9} 
\usepackage{microtype}
\usepackage{setspace} 
\usepackage[utf8]{inputenc}
\usepackage[english]{babel}
\usepackage{times}
\usepackage{array,amsmath}
\usepackage{soul}
\usepackage{cite}
\usepackage[numbers,sort&compress]{natbib}
\usepackage{natbib}

\usepackage{titlesec} 
\titleformat {\section} [block] {\raggedright \fontsize{10}{10}\selectfont\bfseries} {\thesection. \space} {0pt} {}
\titlespacing {\section} {0pt} {12pt} {6pt}
\titleformat {\subsection} [block] {\raggedright \fontsize{10}{10}\selectfont\itshape} {\thesubsection .\space} {0pt} {}
\titlespacing {\subsection} {0pt} {12pt} {6pt}
\titleformat {\subsubsection} [block] {\raggedright \fontsize{10}{10}\selectfont} {\thesubsubsection .\space} {0pt} {}
\titlespacing {\subsubsection} {0pt} {12pt} {6pt}
\titleformat {\paragraph} [block] {\raggedright \fontsize{10}{10}\selectfont} {} {0pt} {}
\titlespacing {\paragraph} {0pt} {12pt} {6pt}

\usepackage{array} \newcommand{\PreserveBackslash}[1]{\let\temp=\\#1\let\\=\temp}
\newcolumntype{C}[1]{>{\PreserveBackslash\centering}m{#1}}
\newcolumntype{R}[1]{>{\PreserveBackslash\raggedleft}m{#1}}
\newcolumntype{L}[1]{>{\PreserveBackslash\raggedright}m{#1}}
\usepackage{lineno}
\usepackage{tabularx}
\usepackage{colortbl}
\usepackage{graphicx}
\usepackage{float}
\usepackage[export]{adjustbox}
\usepackage{caption}
\usepackage{fancyhdr} 
\usepackage{lastpage}
\usepackage{layout}
\usepackage{setspace} 
\usepackage{enumitem}
\usepackage{booktabs}
\usepackage{arydshln}
\usepackage{multirow}
\usepackage{color}
\usepackage{hyperref} 
\hypersetup{
	colorlinks=true,
	linkcolor=blue,
	filecolor=blue,
	urlcolor=black,
	citecolor=cyan,
}

\fancypagestyle{firstpage}{
    \setlength{\headsep}{2.2cm}
    
    \setlength{\footskip}{1.5cm}
    \fancyhf{}
    \lhead{\begin{table}[H]
        \centering
        \begin{tabular}{L{2.5cm}C{10cm}C{3.1cm}R{2cm}}
            \includegraphics[scale=0.035]{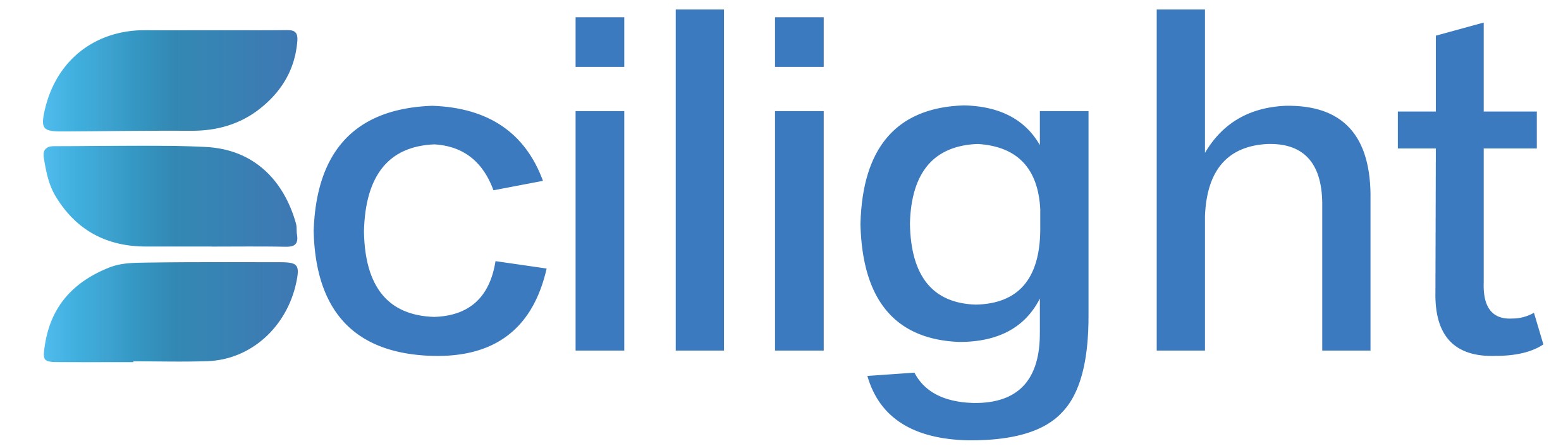} \vspace{-6pt}& \cellcolor{graycolor}\begin{tabular}[c]{@{}c@{}}\textit{International Journal of Gravitation and Theoretical Physics}\\ \href{https://www.sciltp.com/journals/ijgtp}{https://www.sciltp.com/journals/ijgtp}\end{tabular} & \includegraphics[scale=0.012]{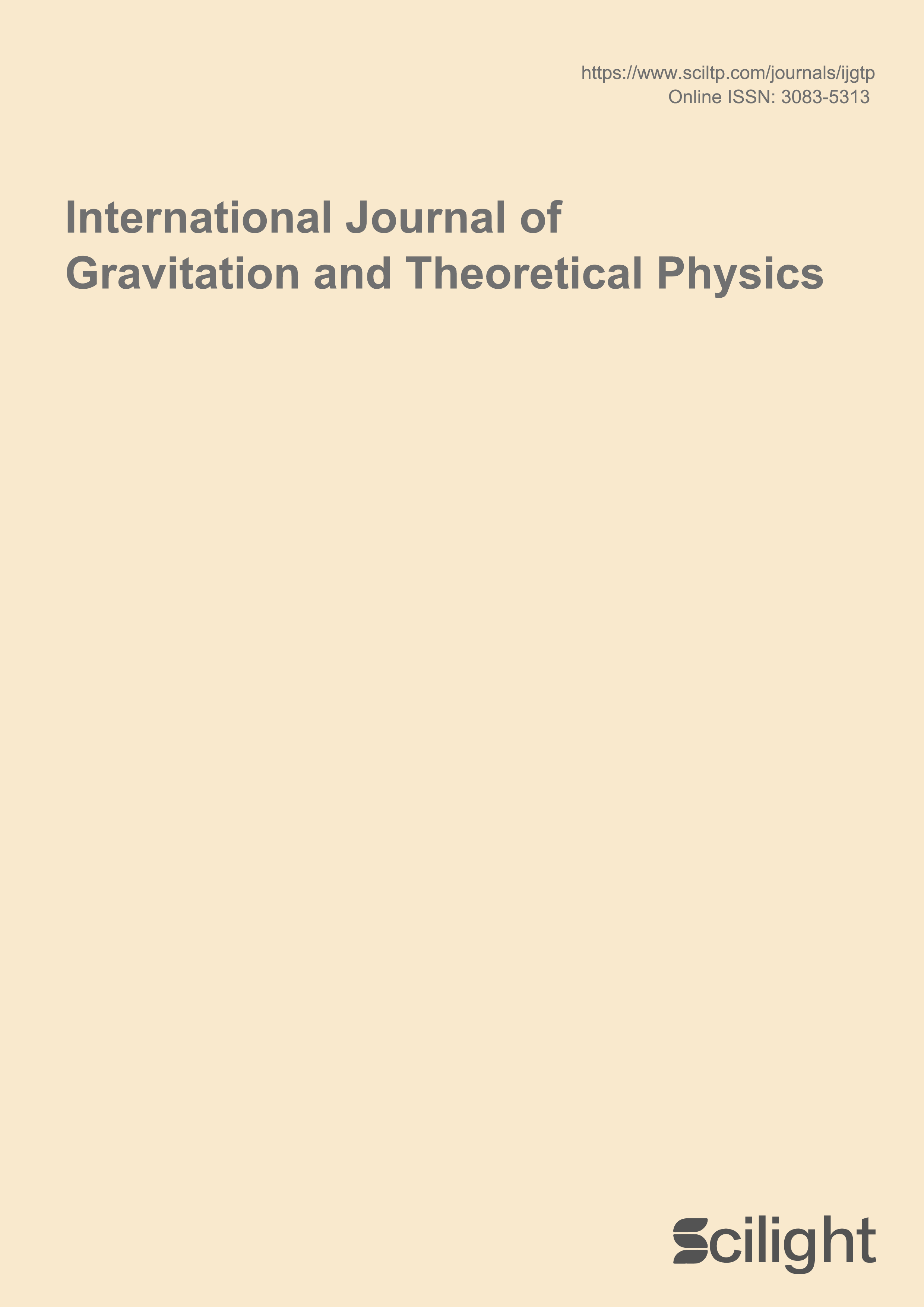} \vspace{-3pt}\\
        \end{tabular}
        \vspace{-22pt}
    \end{table}}
   
    \fancyfoot[C]{
        \vspace{-1.55cm}
        \begin{table}[H]
            \begin{minipage}[c]{0.15\columnwidth}
                \includegraphics[scale=0.5]{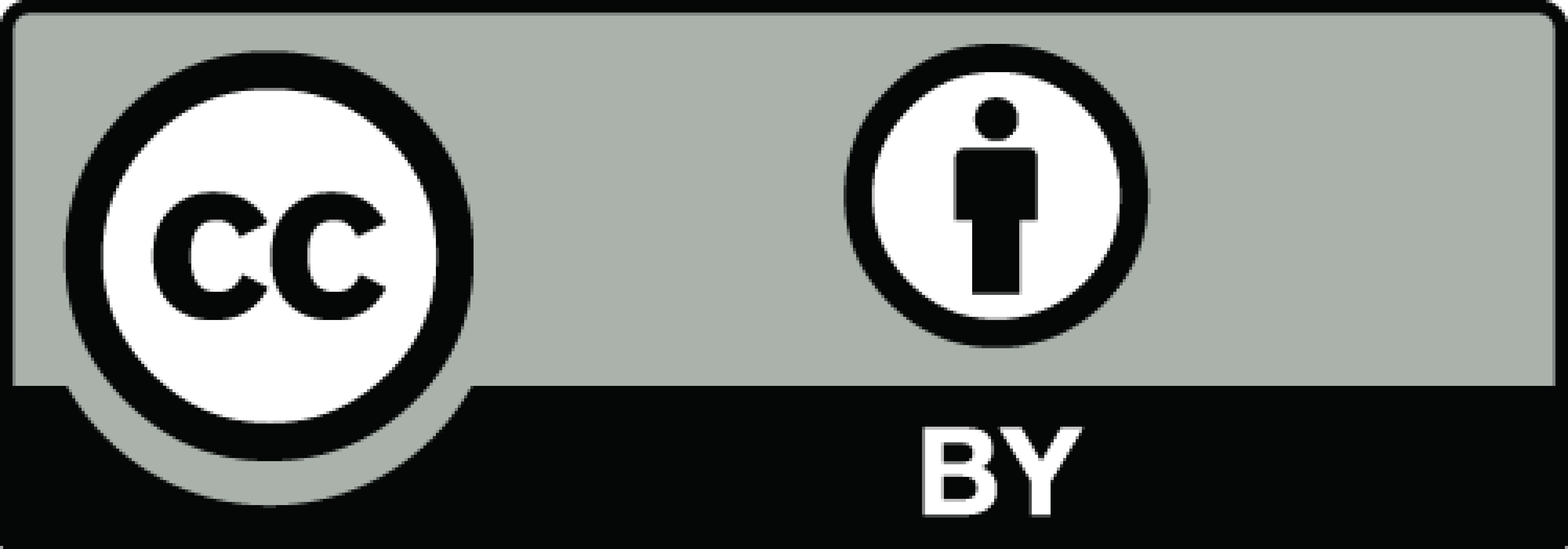} \vspace{1.1pt}
            \end{minipage}
            \hfill
            \begin{minipage}[c]{0.85\columnwidth}
                \scriptsize \textbf{Copyright:} © 2026 by the authors. This is an open access article under the terms and conditions of the Creative Commons Attribution (\mbox{CC BY}) license  (\href{https://creativecommons.org/licenses/by/4.0/}{https://creativecommons.org/licenses/by/4.0/}). \\ \textbf{Publisher’s Note:} Scilight stays neutral with regard to jurisdictional claims in published maps and institutional affiliations.
            \end{minipage}
    \end{table}}
    \vspace{-0.55cm}
}
\def\Order#1{{\cal O}\left(#1\right)}
\setcitestyle{open={[},close={]},citesep={,\!},numbers}

\usepackage[none]{hyphenat}
\begin{document}
\newgeometry{left=2.5cm, right=2.5cm, top=1.8cm, bottom=4cm}
	\thispagestyle{firstpage}
	\nolinenumbers 
	{\noindent \textit{Article}}
	\vspace{4pt} \\
	{\fontsize{18pt}{10pt}\textbf{Hernquist Distribution of Matter as a Source of \\ Black-Hole~Geometry}  }
	\vspace{16pt} \\
	{\large Erdinç Ulaş Saka}
	\vspace{6pt}
	 \begin{spacing}{0.9}
		{\noindent \small
			Department of Physics, Faculty of Science, Istanbul University, Vezneciler, 34134 Istanbul, Türkiye; ulassaka@istanbul.edu.tr
\vspace{6pt}\\
		\footnotesize	\textbf{How To Cite}: Saka, E.U. Hernquist Distribution of Matter as a Source of Black-Hole Geometry. \emph{International Journal of Gravitation and Theoretical Physics} \textbf{2026}, \emph{2}(1), 7. \href{https://doi.org/10.53941/ijgtp.2026.100007}{https://doi.org/10.53941/ijgtp.2026.100007.}}\,\,\href{https://doi.org/10.48550/arXiv.2602.18877}{arXiv:2602.18877 [gr-qc]}\\
	\end{spacing}

\begin{table}[H]
\noindent\rule[0.15\baselineskip]{\textwidth}{0.5pt} 
\begin{tabular}{lp{12cm}}  
 \small 
  \begin{tabular}[t]{@{}l@{}} 
  \footnotesize  Received: day month year \\
  \footnotesize  Revised: day month year \\
   \footnotesize Accepted: day month year \\
  \footnotesize  Published: day month year
  \end{tabular} &
  \textbf{Abstract:} It was recently demonstrated that imposing the condition $P_{r} = -\rho$ on the radial pressure of a galactic halo can lead to regular black-hole solutions for certain density profiles, such as the Dehnen and Einasto models. In the present work, we show that some of the most commonly used halo profiles, including the Hernquist model, do not yield regular geometries under the same condition, but instead support black-hole solutions that retain a central singularity. \\
\\
  & 
\\\\&
\end{tabular}
\noindent\rule[0.15\baselineskip]{\textwidth}{0.5pt} 
\end{table}

\section{Introduction}

Astrophysical black holes are not realized in isolation but are embedded in galactic environments whose large-scale dynamics are dominated by dark matter. A wide range of astronomical observations---from galactic rotation curves to gravitational lensing and large-scale structure formation---provides compelling evidence for the existence of extended dark matter halos surrounding galaxies and clusters \cite{Navarro:1996gj,Bertone:2005xz}. Although the fundamental particle nature of dark matter remains unknown, its macroscopic gravitational influence is well established. In practice, this influence is modeled through phenomenological density profiles, such as the Hernquist and Navarro–Frenk–White (NFW) distributions \cite{Hernquist:1990be,Navarro:1996gj}, which successfully reproduce observed mass distributions in galaxies and are routinely employed in both galactic dynamics and cosmological simulations.

These halo models specify the radial dependence of the energy density but do not uniquely determine the associated pressure components. As a consequence, there is considerable freedom in constructing effective anisotropic fluid descriptions consistent with a given density profile. This freedom allows one to explore different gravitational configurations supported by the same underlying halo phenomenology \cite{Cardoso:2021wlq,Konoplya:2025nqv,Lutfuoglu:2025mqa,Konoplya:2022hbl,Bolokhov:2025zva,Bolokhov:2025fto,Yue:2026ish,Yue:2026evf}.

It has been demonstrated that Dehnen- and Einasto-type density distributions, when supplemented by the condition on the radial pressure $P_{r} = -\rho$, can give rise to regular black-hole geometries free of curvature singularities \cite{Konoplya:2025ect,Zaslavskii:2025oli}. The existence of regular black-hole solutions for this equation of state was first noted in~\cite{Dymnikova:1992ux}, albeit in a setting not associated with specific matter distributions or galactic environments. Motivated by this observation, we investigate the possibility of constructing exact black-hole solutions sourced by realistic dark matter halos, concentrating in particular on the Hernquist profile. We show that, in contrast to certain other density models that naturally produce regular interiors under suitable pressure prescriptions, the Hernquist distribution generically leads to spherically symmetric black-hole solutions that retain a central singularity. Since the Hernquist profile constrains only the density while leaving the pressure unspecified, the adopted pressure choice remains compatible with halo phenomenology, yet it does not automatically guarantee regularity at the center. 

The structure of the paper is as follows. In Section~\ref{sec:solution}, we formulate the general framework for static, spherically symmetric matter configurations and derive the appropriate Einstein field equations governing the geometry. Section~\ref{sec:metrics} is devoted to explicit realizations of the solutions obtained for various parameter choices within the Hernquist density profile, where we discuss their main geometric and physical characteristics. Finally, in Section~\ref{sec:conclusions} we discuss our results.


\section{Spherically Symmetric Dark-Matter Halos and Black-Hole Configurations}
\label{sec:solution}

Dark matter halos surrounding galaxies are commonly described within an effective anisotropic fluid framework, leading to approximately spherical mass distributions on large scales \cite{Benson:2010de}. Various empirical density profiles are used in galactic modeling, depending on the morphology and mass of the host galaxy. A broad class of halo distributions may be represented by the generalized form \cite{Dehnen:1993uh,Taylor:2002zd}
\begin{equation}
\rho(r)
=
2^{(\gamma-\alpha)/k}\,\rho_a
\left(\frac{r}{a}\right)^{-\alpha}
\left(1+\left(\frac{r}{a}\right)^{k}\right)^{-(\gamma-\alpha)/k},
\label{density}
\end{equation}
where $a$ sets the characteristic halo scale and $\rho_a=\rho(a)$ determines the overall normalization. The total mass enclosed within a halo of radius $s>a$ is given by
\begin{equation}
M_{\mbox{halo}} = 4\pi \int_{0}^{s} r^{2}\rho(r)\,dr,
\label{totalhalomass}
\end{equation}
and we assume $\alpha<3$ to guarantee convergence of the integral. Here we will follow the general approach of \cite{Konoplya:2025ect} which will be slightly modified in the end of the section via the introduction of a new constant of integration. 

\subsection{Geometry and Field Equations}

To construct black-hole solutions embedded in such halos, we adopt the general static, spherically symmetric line element
\begin{equation}
ds^{2}
=
- f(r)\,dt^{2}
+
\frac{B^{2}(r)}{f(r)}\,dr^{2}
+
r^{2}\left(d\theta^{2}+\sin^{2}\theta\,d\varphi^{2}\right),
\label{line-element}
\end{equation}
with $B(r)>0$. 

Instead of introducing $f(r)$ directly, it is convenient to define the Misner--Sharp mass function $m(r)$ through
\begin{equation}
\frac{B^2(r)}{f(r)}
=
\left(1 - \frac{2m(r)}{r}\right)^{-1}.
\label{massfunction}
\end{equation}

The event horizon $r_0$ is located at the largest root of $f(r_0)=0$, which is equivalent to
\begin{equation}
r_0 = 2m(r_0).
\label{horizoncond}
\end{equation}

The matter content is described by an anisotropic stress-energy tensor of diagonal form,
\begin{equation}
T^{\mu}{}_{\nu}
=
\mathrm{diag}
\bigl(-8\pi\rho,\; 8\pi P_r,\; 8\pi P,\; 8\pi P\bigr),
\label{stress-energy}
\end{equation}
where $\rho(r)$ is the energy density, $P_r(r)$ the radial pressure, and $P(r)$ the tangential pressure.

Einstein's equations reduce to two independent differential relations,
\begin{eqnarray}
\frac{dm}{dr}
&=&
4\pi r^{2}\rho(r),
\label{mder}\\
\frac{dB}{dr}
&=&
4\pi r^{2}B(r)\,
\frac{\rho(r)+P_r(r)}{r-2m(r)}.
\label{Bder}
\end{eqnarray}

Finally, using (\ref{massfunction}), we obtain the expression for $f(r)$ as follows:
\begin{equation}
    f(r)=B(r)^2\left(1 - \frac{2m(r)}{r}\right).
\end{equation}

\subsection{Horizon Regularity and Equation of State}

Regularity of $B(r)$ at the horizon requires that the numerator in Equation~(\ref{Bder}) vanish when $r\to r_0$, since $r-2m(r)$ tends to zero there. Consequently,
\begin{equation}
\rho(r_0)+P_r(r_0)=0.
\label{horizonregular}
\end{equation}

Following \cite{Konoplya:2025ect}, we
adopt the vacuum-like equation of state
\begin{equation}
P_r(r) = -\rho(r),
\label{vacuum}
\end{equation}
which automatically satisfies condition~(\ref{horizonregular}).

It is important to stress that empirical halo profiles such as the Hernquist model prescribe only $\rho(r)$, without specifying a unique relation between pressure and density. In collisionless descriptions based on the Jeans equations, pressure emerges effectively from velocity dispersion and is not constrained by a simple equation of state. Therefore, choosing $P_r=-\rho$ does not contradict the phenomenological foundations of the Hernquist profile; rather, it represents a consistent effective description that ensures desirable properties of the resulting spacetime geometry.

\subsection{Explicit Form of the Solution}

With the condition (\ref{vacuum}), Equation~(\ref{Bder}) integrates trivially to
\begin{equation}
B(r)=1,
\label{Bsol}
\end{equation}
so that the metric simplifies to the standard Schwarzschild-like gauge.

The tangential pressure then follows from the angular components of Einstein's equations,
\begin{equation}
P(r)
=
-\rho(r)
-\frac{r}{2}\frac{d\rho}{dr}.
\label{Pex}
\end{equation}

For the generalized density profile (\ref{density}), this can be written explicitly as
\begin{equation}
P(r)
=
-\frac{\rho(r)}{2}\,
\frac{a^{k}(2-\alpha)+r^{k}(2-\gamma)}
     {a^{k}+r^{k}}.
\end{equation}

Finally, integrating Equation~(\ref{mder}) gives
\begin{equation}
m(r)
=
4\pi \int_{0}^{r} x^{2}\rho(x)\,dx
+
M_{0},
\label{msol}
\end{equation}
where $M_{0}$ is an integration constant associated with the central mass contribution. In \cite{Konoplya:2025ect} this constant is equal to zero, which is not compulsory in our case. 

Thus, starting from a prescribed halo density profile and adopting a physically admissible radial equation of state, one obtains a family of exact spherically symmetric black-hole solutions supported by dark matter.

\section{Illustrative Families Of Solutions}
\label{sec:metrics}

Of particular interest are halo configurations with $\gamma>3$, for which the total mass integral converges as $s\rightarrow\infty$. In this case, the dark matter distribution extends to arbitrarily large radii while still producing a finite asymptotic mass. This property allows one to construct analytically tractable black-hole geometries supported by matter that is present everywhere in space.

\begin{figure}[H]
\centering
\includegraphics[scale=1.0]{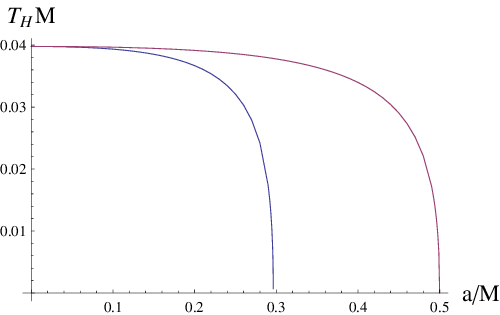}
\caption{Hawking temperature as a function of halo parameter $a$ for the two models: Hernquist model $\alpha=1$, $\gamma=4$, $k=1$, $M_{0}=0$ (red) and $\alpha=0$, $\gamma=4$, $k=1$, $M_{0}=0$ (blue).}\label{fig:T}\hl{} 
\end{figure}

\subsection{Hernquist Profile: $\alpha=1$, $\gamma=4$, $k=1$}

The Hernquist model \cite{Hernquist:1990be} corresponds to the parameter choice $\alpha=1$, $\gamma=4$, $k=1$. For this case, the mass function can be integrated explicitly, leading to
\begin{equation}
f(r)
=
1-\frac{2m(r)}{r}
=
1-\frac{2M_0}{r}
-\frac{2(M-M_0)\,r}{(r+a)^2}.
\label{pureHernquist}
\end{equation}
The total asymptotic mass is obtained from
\begin{equation}
M
=
4\pi\int_{0}^{\infty}\rho(r)\,r^{2}dr
+
M_0
=
16\pi\rho_a a^{3}
+
M_0.
\label{MHernquist}
\end{equation}

The solution possesses a singularity in the center for any $a$, $M_0$ and $M$, since the Kretschmann scalar diverges as $r\to0$,
$$\begin{array}{rcl}
R_{\mu\nu\lambda\sigma}R^{\mu\nu\lambda\sigma}&=&\dfrac{48 M_0^2}{r^6}+\dfrac{32(M-M_0)(M+2M_0)}{a^4 r^2}\\
&&+\dfrac{192 \left(M_0^2-M^2\right)}{a^5 r}+\Order{1}.
\end{array}$$

If $M_0>0$, the geometry generically admits an event horizon. In the limiting case $M_0=0$, the existence of a horizon depends on the halo scale parameter: a black hole forms provided $0<a\le M/2$.

Hawking temperature is given by the relation
\begin{equation}\label{metric}
  T_H=\frac{f'(r)}{4\pi}\Biggr|_{r=r_0},
\end{equation}
where $r_0$ is the radius of the event horizon.
The Hawking temperature in the limit $M_0=0$ has a simple form,
\begin{eqnarray}
T_H &=& \frac{M \left(\sqrt{M (M-2 a)}-2 a+M\right)}{2 \pi 
   \left(\sqrt{M (M-2 a)}+M\right)^3} ,
\\\nonumber
  &=&\frac{1}{8 \pi  M}-\frac{a^2}{32 \pi M^3}+\Order{a}^3.
\end{eqnarray}

Thus, we see that the galactic halo leads to only a slight decrease of the Hawking temperature (see Figure~\ref{fig:T}).

\subsection{Generalized Profile with $k=2$}

A natural extension is obtained by keeping $\alpha=1$, $\gamma=4$ while choosing $k=2$. The resulting metric function takes the form
\begin{equation}
f(r)
=
1-\frac{2M_0}{r}
-\frac{2(M-M_0)\,r}
     {\bigl(\sqrt{a^{2}+r^{2}}+a\bigr)\sqrt{a^{2}+r^{2}}},
\label{k2blackhole}
\end{equation}
and the corresponding asymptotic mass reads
\begin{equation}
M
=
4\pi\int_{0}^{\infty}\rho(r)\,r^{2}dr
+
M_0
=
8\sqrt{2}\,\pi\rho_a a^{3}
+
M_0.
\label{Mk2}
\end{equation}

In this configuration, an event horizon exists for any $a>0$ provided $M>M_0\ge0$. Thus, the parameter range leading to black-hole solutions is broader than in the $k=1$ Hernquist case.
The Kretschmann scalar again diverges for any values of the parameters,
$$
R_{\mu\nu\lambda\sigma}R^{\mu\nu\lambda\sigma}=\dfrac{48 M_0^2}{r^6}+\dfrac{4(M-M_0)(2M-5M_0)}{a^4 r^2}+\Order{1}.
$$

\subsection{Case $\alpha=2$, $\gamma=4$, $k=1$}

Choosing $\alpha=2$ with $\gamma=4$ and $k=1$ leads to a particularly simple analytic expression:
\begin{equation}
f(r)
=
1-\frac{2M_0}{r}
-\frac{2(M-M_0)}{r+a}.
\label{a2blackhole}
\end{equation}

The total mass is again
\begin{equation}
M
=
16\pi\rho_a a^{3}
+
M_0.
\end{equation}

As in the previous cases, $M_0>0$ guarantees the existence of a horizon. When $M_0=0$, a black-hole solution is obtained if the halo parameter satisfies $0<a\le M/2$. The central singularity persists for any values of the parameters, because
$$
R_{\mu\nu\lambda\sigma}R^{\mu\nu\lambda\sigma}=\dfrac{48 M_0^2}{r^6}+\dfrac{32(M-M_0)M_0}{a r^5}+\Order{\frac{1}{r^4}}.
$$

The Hawking temperature is 
\begin{equation}
T_H = \frac{1}{8 \pi  M}+\frac{a^2 M_{0} (M-M_{0})}{32
   \pi  M^5}+\Order{a}^3,
\end{equation}
which again means relatively small correction to the temperature due to halo. When $M_{0}=0$, the Hawking temperature does not depend on $a$, coinciding with that of the Schwarzschild black hole.

\subsection{Regular Configuration: $\alpha=0$, $\gamma=4$, $k=1$}

Finally, consider the case $\alpha=0$, $\gamma=4$, $k=1$. Here both the density and the pressure remain finite at the center, and therefore the geometry may be regular if the central mass parameter vanishes. The lapse function~becomes
\begin{equation}
f(r)
=
1-\frac{2M_0}{r}
-\frac{2(M-M_0)\,r^{2}}{(r+a)^{3}},
\label{a0blackhole}
\end{equation}
with the asymptotic mass
\begin{equation}
M
=
\frac{64}{3}\pi\rho_a a^{3}
+
M_0.
\end{equation}

The Kretschmann scalar diverges only if $M_0\neq0$,
$$\begin{array}{rcl}
R_{\mu\nu\lambda\sigma}R^{\mu\nu\lambda\sigma}&=&\dfrac{48 M_0^2}{r^6}+\dfrac{96 (M-M_0)M_0(6r-a)}{a^5 r^2}\\&&\\
&&+\dfrac{96 (M-M_0)(M-21M_0)}{a^6}+\Order{r}.
\end{array}$$

When $M_0=0$, the Kretschmann scalar if finite everywhere, the solution has a de Sitter core and coincides with the regular black-hole geometry previously obtained in \cite{Konoplya:2025ect}. In contrast to the Hernquist-type profiles discussed above, the absence of a central mass term here allows for a nonsingular interior supported entirely by the halo~matter.

These examples illustrate how different parameter choices within the same general density framework lead to qualitatively distinct black-hole geometries. Depending on the interplay between the halo scale $a$ and the central mass parameter $M_0$, one may obtain either singular or regular solutions, as well as different horizon structures.

\section{Discussions and Conclusions}
\label{sec:conclusions}

Black holes embedded in galactic dark-matter halos have been the subject of numerous investigations in recent years. For singular geometries surrounded by realistic halo distributions, a broad range of physical phenomena has already been explored. These include the analysis of quasinormal-mode spectra and ringdown behavior \cite{Zhao:2023tyo,Daghigh:2022pcr,Zhang:2021bdr,Konoplya:2021ube,Konoplya:2022hbl,Dubinsky:2025fwv,Feng:2025iao,Pezzella:2024tkf,Chakraborty:2024gcr,Liu:2024bfj,Liu:2024xcd}, the computation of grey-body factors and absorption properties \cite{Mollicone:2024lxy,Hamil:2025pte}, as well as optical signatures such as shadows and gravitational lensing patterns \cite{Kouniatalis:2025itj,Fernandes:2025osu,Chen:2024lpd,Konoplya:2025nqv,Hou:2018avu,Tan:2024hzw,Figueiredo:2023gas,Konoplya:2025mvj,Macedo:2024qky,Xavier:2023exm}. In addition, dynamical aspects---most notably accretion flows and related astrophysical processes---have also been addressed within this framework \cite{Chowdhury:2025tpt,Heydari-Fard:2024wgu}.

In this work we have constructed and analyzed a class of static, spherically symmetric black-hole solutions sourced by galactic dark matter halos within the framework of Einstein gravity. Starting from a generic density profile describing realistic halo distributions, we showed that imposing the vacuum-like radial equation of state $P_r=-\rho$ leads to exact analytic solutions. 
This condition guarantees regular behavior of the metric functions at the event horizon without altering the prescribed density profile, but it does not by itself ensure regularity at the spacetime center.

We investigated several representative choices of halo parameters, including the widely used Hernquist profile. Our analysis demonstrates that, in contrast to certain Dehnen- and Einasto-type models, the Hernquist distribution does not generically support a regular black-hole interior under the same pressure condition. Instead, the resulting geometries typically retain a central singularity whenever a nonvanishing integration constant $M_0$ is present. Only in special cases, such as profiles with finite central density (e.g., $\alpha=0$), can fully regular black-hole solutions be obtained when the central mass term vanishes \cite{Konoplya:2025ect}.

We also identified the parameter domains in which event horizons form and clarified how the interplay between the halo scale $a$ and the integration constant $M_0$ determines the global structure of the spacetime. The solutions constructed here provide explicit examples of black holes embedded in realistic galactic environments and illustrate how different halo profiles can lead to qualitatively distinct interior geometries.

Our results highlight that the presence of dark matter alone does not automatically guarantee regularization of the central singularity; rather, the outcome crucially depends on the detailed structure of the density profile and the chosen effective pressure prescription. These findings contribute to a better understanding of how astrophysically motivated matter distributions influence black-hole geometries in general relativity.
In this context analysis of quasinormal modes for the obtained solution in numerical \cite{Leaver:1985ax,Gundlach:1993tp,Fortuna:2020obg,Dubinsky:2024gwo} and analytic \cite{Dubinsky:2024rvf,Malik:2024voy} ways could be possible direction of further research.
	

%
%
 
        \section*{Institutional Review Board Statement}
Not applicable.

		\section*{Informed Consent Statement}
Not applicable.

		\section*{Data Availability Statement}
Not applicable.
 

		\section*{Conflicts of Interest}
The authors declare no conflict of interest.

\section*{Use of AI and AI-Assisted Technologies}
No AI tools were utilized for this paper.

	\small
	\bibliographystyle{scilight}
	
	

\end{document}